# Ellipsometric detection of transitional surface structures on decapped GaAs(001)

**[&]A.V. Vasev[1], S.I. Chikichev[1,2]**

[1]Institute of Semiconductor Physics, Siberian Branch of the Russian Academy of Sciences, Laboratory of Optical Ellipsometry, Acad. Lavrentiev Avenue, 13, 630090, Novosibirsk, Russia
[2]Novosibirsk State University, Faculty of Physics, Department of Semiconductor Physics, Pirogova street, 2, 630090, Novosibirsk, Russia
[&]Corresponding author: **email:** vasev@isp.nsc.ru; **tel./fax:** 7 (383-2) 333-884;

**Abstract.** Structural and optical properties of MBE-grown GaAs(001) surface have been studied by reflection high-energy electron diffraction and single-wavelength ellipsometry under dynamic conditions of ramp heating after desorption of passivating As-cap-layer with and without $As_4$ beam applied to the surface. For a number metastable reconstruction transitions a clear correlation is established between diffraction and optical data. Boundary lines for transitional superstructures are determined as a function of As flux and corresponding activation energies are estimated. For the first time it is shown ellipsometrically that optical response of the surface is drastically different for transitions of the *order➛order* and *order➛disorder* type.

## Introduction

A technologically important GaAs(001) surface is known to exhibit a large variety of surface superstructures depending on preparation conditions [1]. Thermodynamically stable superstructures $c(4\times4)$, As-rich $(2\times4)$ and Ga-rich $(4\times2)$ are most relevant to molecular beam epitaxy of GaAs(001) and they have been thoroughly studied using a number of surface science techniques. Much less is known about the group of co-called transitional reconstructions with apparent symmetries $(2\times3)$, $(3\times1)$, $(1\times6)$, $(4\times6)$ etc, which can be prepared on the GaAs(001) surface using (sometimes very) specific procedures. The purpose of the present work is to show that simple single-wavelength ellipsometry is quite useful for dynamical monitoring of surface structure transitions including those between metastable ones.

Ellipsometry is an optical technique designed to determine accurately the ratio of complex reflectances $R_p$ and $R_s$ from the sample, where $R_p$ and $R_s$ are the reflection coefficients of light polarized parallel to (p) or perpendicular to (s) the plane of incidence. The ratio is usually written as $\tan(\psi)e^{i\Delta} = R_p/R_s$, where the values of $\psi$ and $\Delta$ characterize amplitude and phase of the complex ratio. The presence of reconstruction on the crystal surface means that atomic structure here is different from the bulk leading to different optical properties of the superficial layer. The $\psi$ and $\Delta$ are very sensitive to changes in the surface conditions including composition, morphology and other parameters of the sample [2,3]. Using simultaneously the reflection high-energy electron diffraction makes it possible to correlate structural and optical properties of the crystal surface.

## Experimental

Investigations of GaAs(001) surface optical and structural properties were performed in MBE-type UHV chamber, equipped by single-wavelength ellipsometer, electron gun for reflection high-energy electron diffraction (RHEED) and arsenic molecular beam source. A He-Ne laser was used as a light source ($\lambda$=6328 Å). Ellipsometric parameters $\psi$ and $\Delta$ were measured with an accuracy of 0.01° and 0.1°, respectively. The angle of incidence of light beam on the sample surface was fixed at 71.95±0.05°. RHEED pictures were taken at electron energy of 17 keV. Base pressure in the chamber was less than $4\times10^{-9}$ torr and did not exceed $5\times10^{-8}$ torr at the highest substrate tem-perature. Beam equivalent pressure (BEP) of the As$_4$ at the sample surface was measured by nude ion gauge and varied in our experiments from $5\times10^{-8}$ torr up to $3\times10^{-5}$ torr. Experimental setup used in the present work is shown schematically in Fig.1.

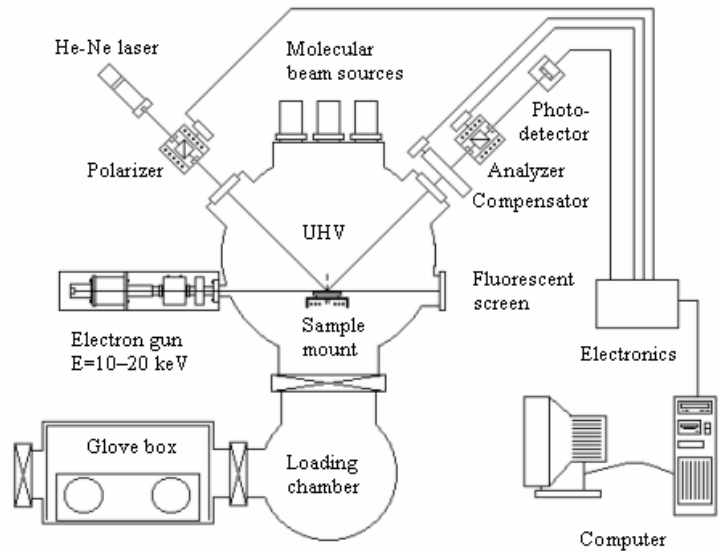

**Fig.1.** Schematical representation of experimental setup.

The samples used were homoepitaxial layers of GaAs (~70 nm thick) grown by molecular-beam epitaxy on semi-insulating substrates oriented (001) within ±0.5° and capped by amorphous As layer (~10 nm thick) immediately after growth using As$_2$ beam from cracker-cell. After air transfer to our chamber the clean surface was obtained by removing the As-cap-layer by heating the sample in ultra-high vacuum (UHV). We used ramp heating with a rate of 2°C/min for the first cycle and ±8°C/min if further heating (cooling) were required.

Sample temperature was measured by chromel-alumel thermocouple, which contacted the molybdenum carrier from the back side. Molten indium was used to fix the sample on the front side of the Mo carrier and provide a good thermal contact with it. In a separate experiment we attached additional (external) thermocouple to the front surface of the Mo carrier and compare the readings of two thermocouples. Results are shown in Fig.2. It can be seen that for temperatures higher than 550°C the readings of external thermocouple are systematically lower, the difference being as high as 25°C at ~650°C. When temperature is increased through the melting point of indium small changes of sample position usually occur, and this is easily detected by ellipsometer. The insert in Fig.2 shows the behavior of ψ angle during sample heating from 20°C up to 180°C. At the In melting point a smooth curve is disrupted due to slight change in the incidence angle. We used this effect for absolute calibration of both thermocouples. The corresponding point is indicated by star (☆) on the curve. This effect can also be used as a sensitive indicator of how good is thermal contact between internal thermocouple and sample carrier.

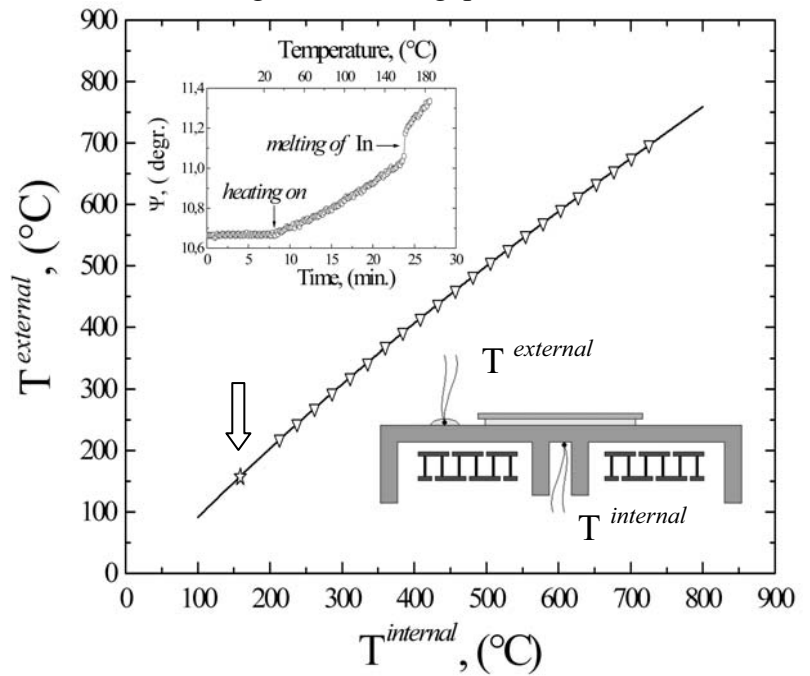

**Fig.2.** Calibration curve for thermocouple imbedded into molybdenum sample carrier. The insert shows the dependence of ellipsometric ψ parameter (which is most sensitive to changes of the incidence angle) on temperature. A jump-like discontinuity is clearly seen at the In melting point.

**Results and discussion**

The evolution of ellipsometric parameters ψ and Δ during heating of decapped samples is shown in Fig.3 for different As$_4$ beam intensities applied to the surface. All curves look quite similar shifting to higher temperature with increasing As$_4$ BEP. Such a behavior of optical properties indicates that features observed are surface-related. Comparison of ellipsometric data with RHEED patterns observed in specific temperature ranges shows clear correlation between measured optical response and surface superstructures on decapped GaAs(001).

To illustrate this point in more detail let us consider the results obtained for BEP(As$_4$)=1.8×10$^{-7}$ torr, which are shown enlarged in Fig.4. After sublimation of As-cap layer the following sequence of surface structures was observed by RHEED: (2×3)→γ(2×4)→

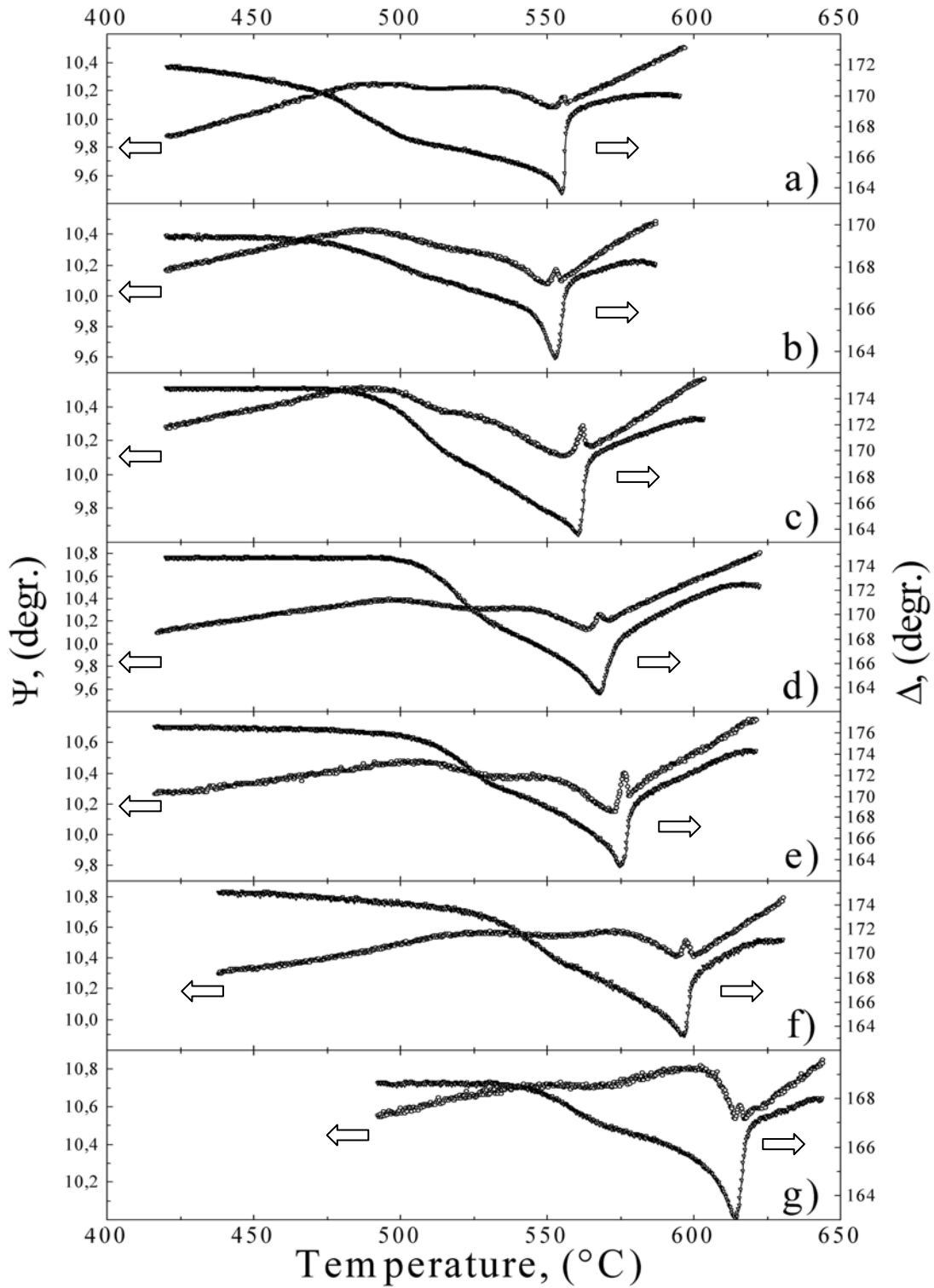

**Fig.3.** Evolution of ellipsometric parameters ψ and Δ during decapped GaAs(001) heating for different As$_4$ beam equivalent pressures (in torr):
a) 1.6×10$^{-8}$; b) 4×10$^{-8}$; c) 1.8×10$^{-7}$; d) 6.5×10$^{-7}$; e) 2.3×10$^{-6}$; f) 7.5×10$^{-6}$; g) 2.8×10$^{-5}$.

β(2×4)→(3×1)→(1×6)→(4×6) →(4×2) as shown in Fig.5. Each of these superstructures exists within a particular temperature range. The boundaries of the superstructure "existence" regions (indicated by arrows in Fig.4) coincide remarkably well with those points on ψ($T_s$) and Δ($T_s$) curves where temperature derivatives are changed markedly. Up to substrate temperature of $T_s$=490°C ψ increases with temperature and then start to decrease. Parameter Δ also decreases for $T_s$>490°C. Within a three-phase model (bulk GaAs – reconstructed surface layer – vacuum) this behavior can be interpreted as being due to increase of absorption index in the superficial layer. This appears to be reasonable since transition from As-rich (2×4) reconstruction toward As-poor (3×1) is accompanied by Ga dimer formation on the surface. In the temperature range from 350°C to 490°C no gallium dimers are present on the surface and absorption index remains relatively unchanged.

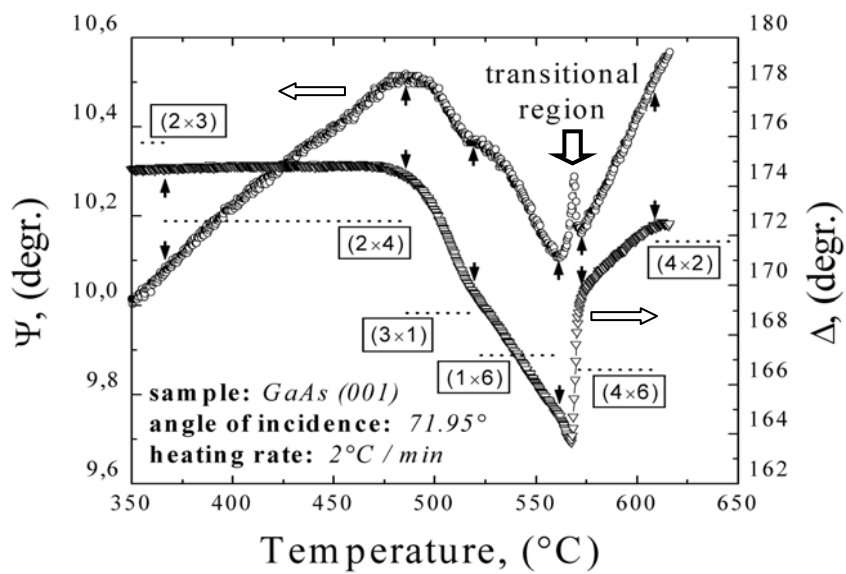

**Fig.4.** Evolution of ellipsometric parameters ψ and Δ during decapped GaAs(001) heating under $As_4$ beam with BEP(As)=1.8×10$^{-7}$ torr. Dotted lines indicate the temperature ranges for different superstructures observed by RHEED.

But most interesting observation concerns the transition between two metastable reconstructions (1×6) and (4×6), which is characterized by a spike on the ψ($T_s$) curve accompanied by sharp rise of Δ($T_s$). Careful RHEED observations within this region have shown in fact that surface disorders dramatically here up to the point that only bulk-like (1×1) structure is observed. Thus, in the narrow temperature range 555–565°C two surface phase transitions seems to occur: (1×6)→(1×1) and (1×1)→(4×6). The first is of the *order→disorder* type, while the second is a reverse. Since all other transitions observed are of the *order→order* type we think that ψ spike and Δ jump is a characteristic feature of two almost overlapping *order↔disorder* transitions. Adatoms with increased number of broken bonds are more abundant on disordered surface and this can be the major reason for substantially altered polarizability in the surface layer, which is registered ellipsometrically.

It should be pointed out that (4×6) is best visible at the lowest $As_4$ pressure employed, becoming dim and poorly resolved with increasing $As_4$ flux to the surface. Apparently the incoming $As_4$ molecules impede long-range ordering on (4×6) surface but, nonetheless, ψ spike was clearly observed even at high $As_4$ flux. This permitted us to track the relevant phase boundary over the large $As_4$ overpressure range.

The reciprocal temperatures corresponding to boundary points of various reconstructions are plotted in Fig.6 for different $As_4$ beam intensities incident on the sample. It is seen that in the pressure range from 2.3×10$^{-6}$ to 4×10$^{-5}$ torr the $As_4$ flux corresponding to superstructure boundary points varies linear with inverse temperature. Activation energies for linear parts of the curves are shown in Fig.6. It can be seen that activation energies for (2×4)→(3×1) and (3×1)→(1×6) transitions are about 1eV lower than for (1×6)→(1×1) and (1×1)→(4×6). Thus, surface disordering requires more energy than for transition to the state of different order. With decreasing $As_4$ flux the boundary points deviate progressively from linear dependence and for BEP($As_4$)<1.8×10$^{-7}$ torr became almost insensitive to the flux.

Not all surface structure transitions can be discriminated by ellipsometry. As a rule, just after removing the As cap-layer the surface exhibits RHEED pattern with apparent (2×3) symmetry, which evolves into so-called γ(2×4) reconstruction during heating. In terms of electron diffraction this reconstruction is characterized by greatly diminished intensity of half-order streak in the [$\bar{1}$10] azimuth [4] (Fig.5 (d)). On further heating the γ(2×4) surface transforms into β(2×4) (Fig.5 (e)), but ψ($T_s$) and Δ($T_s$) curves remain smooth and featureless in this temperature interval. Obviously, optical properties of the surface change very slightly in the course of γ(2×4)→β(2×4) transitions.

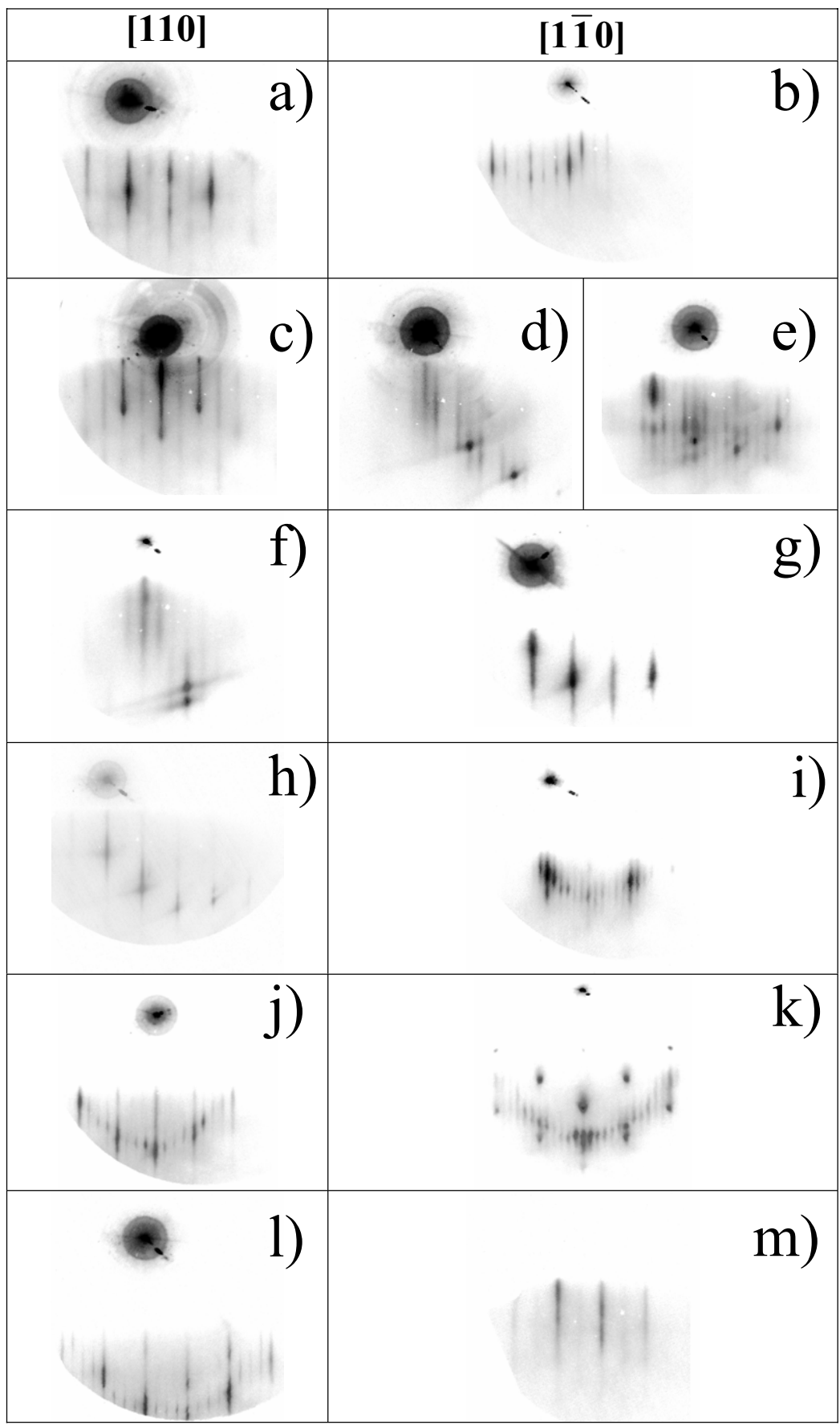

**Fig.5.** RHEED patterns (negatives) of surface structures observed in [110] and [1̄10] azimuths during heating. (a,b) – (2×3); (c,d) – γ(2×4); (c,e) – β(2×4); (f,g) – (3×1); (h,i) – (1×6); (j,k) – (4×6); (l,m) – (4×2).

Finally, the metastability of 6-fold surface structures should be stressed. They are observed only during first heating cycle. If the sample with (4×2) reconstructions is cooled under $As_4$ beam the usual sequence of transitions occur (4×2)→(3×1)→(2×4) in agreement with static surface phase diagram [5].

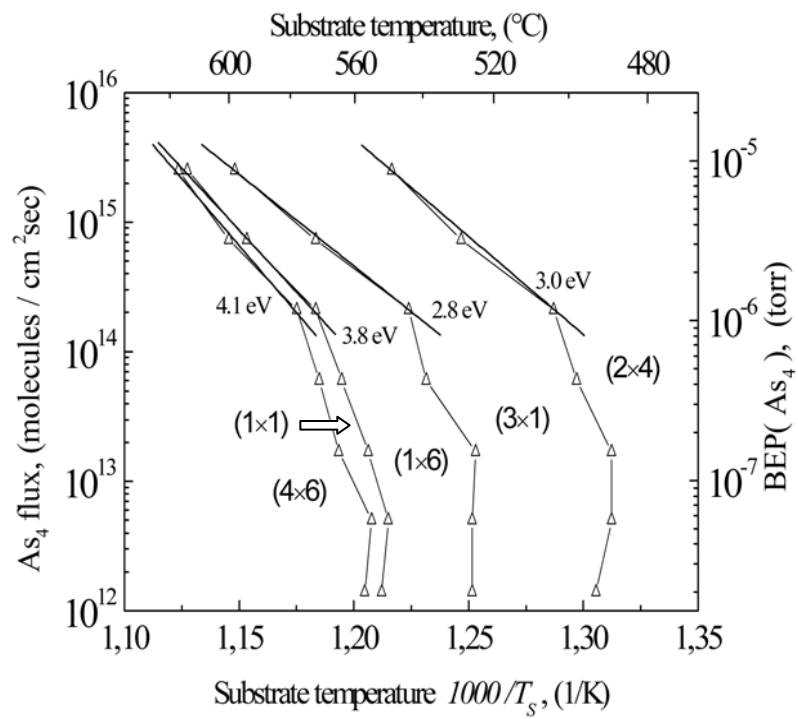

**Fig.6.** Boundary lines for transitional superstructures (3×1), (1×6), (1×1) and (4×6).

## Conclusions

We have demonstrated in the present work that single-wavelength ellipsometry can be used for *in-situ* monitoring of superstructure transitions on decapped GaAs(001) surface under dynamical conditions of continuous heating. Boundary lines for transitional superstructures (3×1), (1×6), (1×1) and (4×6) have been determined as a function of $As_4$ flux and for BEP($As_4$)>2.3×10$^{-6}$ torr activation energies for corresponding transitions have been estimated. For transitions involving disordered surface with apparent symmetry (1×1) the activation energies are about 1eV higher than for transitions of the *order→order* type. Optical response of the surface also differs dramatically for two types of transitions.

## Acknowledgments

The authors are very thankful to Dr. B.R. Semyagin, M.A. Putyato and Dr. V.V. Preobrazhenskii for As-capped samples growth and useful discussions. The present work is partly supported by Ministry of Education of Russian Federation through the Task Program "Integration" (Contract № 0765/785).## References